\documentclass[fleqn,10pt, reprint,' amsmath,amssymb,
 aps]{wlscirep}
\usepackage[utf8]{inputenc}
\usepackage[T1]{fontenc}
\usepackage{graphicx}% Include figure files
\usepackage{caption}
\usepackage{subcaption}
\usepackage{dcolumn}% Align table columns on decimal point %MEXI AQUI
\usepackage{bm}% bold math   MEXI AQUI
\usepackage{booktabs}  % MEXI AQUI
\usepackage{subcaption}  %MEXI AQUI

\title{The Multiplex Efficiency Index: unveiling the  Brazilian Air Transportation Multiplex Network - BATMN}

\author[1,2,*]{Izabela M. Oliveira}
\author[2,+]{Laura C. Carpi}
\author[2,3,4,+]{A.P.F. Atman}
\affil[1]{Departamento de Matem\'atica, Centro Federal de Educa\c c\~ao Tecnol\'ogica de Minas Gerais, CEFET-MG. Av. Amazonas, 7675, Belo Horizonte, MG, Brasil. CEP: 30.510-000.}
\affil[2]{Programa de P\'os-Gradua\c c\~ao em Modelagem Matem\'atica e Computacional, PPGMMC, CEFET-MG.}
\affil[3]{Departamento de F\'isica, CEFET-MG}
\affil[4]{Instituto Nacional de Ci\^encia e Tecnologia de Sistemas Complexos, INCT-SC, CEFET-MG.}

\affil[*]{izabelamarques@cefetmg.br}

\affil[+]{these authors contributed equally to this work}

\begin{abstract}
Modern society is increasingly massively connected, reflecting an omnipresent tendency to organize social, economic, and technological structures in complex networks. Recently, with the advent of the so-called multiplex networks, new concepts and tools were necessary to better understand the characteristics of this type of systems, as well as to analyze and quantify the performance and efficiency of these systems. The concept of diversity in multiplex networks is a striking example of this intrinsically interdisciplinary effort to better understand the nature of complex networks. In this work, we use data related to the Brazilian air transport network to introduce a new index, Multiplex Efficiency Index, which allows quantifying the temporal evolution of connectivity diversity, particularly when the number of layers of the multiplex network varies over time. We demonstrate through the definition of the new index, how the Brazilian air transportation network has being changing over the years due to the privatization processes of airports and mergers of airlines in Brazil. Besides that, we show how the Multiplex Efficiency Index is able to quantify fluctuations in network efficiency in a non-biased way, limiting its values between 0 and 1, taking into account the number of layers in the multiplex structure. We believe that the proposed index is of great value for the evaluation of the performance of any multiplex network, and to analyze, in a quantitative way, its temporal evolution independently of the variation in the number of layers.

\end{abstract}
\begin{document}

%\flushbottom
\maketitle
\thispagestyle{empty}

\section*{\centering{Introduction}}
Efficiency, according to the Cambridge Dictionary is "the condition or fact of producing the results you want without waste, or a particular way in which this is done" or, by Oxford, "the quality of doing something well with no waste of time or money”.
In the context of complex networks, efficiency refers to the navigability of the network, based on the distance from the nodes, so that networks with smaller shortest paths are more efficient~\cite{vragovic2005efficiency, hollingshad2012new, latora2001efficient}. In the case of multiplex networks, a special case of multilayer structure in which the same set of nodes is connected by more than one type of relations in a layered structure~\cite{definMultiplex}, it is shown that structural aspects, such as overlapping edges, number of layers and degree correlations between layers, influence the efficient exploration of the network through random walkers~\cite{battiston2016efficient}.
The concept of efficiency in multiplex networks can be related to the level of diversification of the network paths in the different layers. A greater diversification implies in the presence of less overlapping links between layers, reducing the redundancy of the connections, and expanding the possibilities of using different paths to explore the structure in a more efficient way.

In this sense, supported by the definition of diversity of a multiplex network introduced in a recent work~\cite{carpi2019assessing}, we present the Multiplex Efficiency Index ($\mathcal{E}$-index), which is a modified version of the diversity measure capable of quantifying the connectivity patterns of a multiplex network on a normalized scale from 0 to 1, which is independent of the number of layers. 
 
In this way, an $\mathcal{E}$-index value close to 1 is indicative of a high paths diversification. On the contrary, if $\mathcal{E}$-index is close to 0, connecting paths possess more redundancy.
The $\mathcal{E}$-index is thought to evaluate temporal evolution of multiplex networks when the number of layers changes over time. Therefore, the novelty introduced by this measure is the advantage of allowing the comparison between multiple multiplex networks, regardless of the number of layers present in the structure.

There are numerous possibilities of applications of the $\mathcal{E}$-index in different areas. For example, to quantify connection differences in social networks~\cite{d2014networks}, showing user sets that have similar or distinct connections, to analyze networks of criminal relationships~\cite{toledo2020diversity}, to point out similarities (or differences) between the preferences of a group of people or to verify the protein interactions of the human HIV-1 virus~\cite{carpi2019assessing}, to check the supply of transport modalities for a given region~\cite{aleta2017multilayer}, among many others. 

In this work we apply the $\mathcal{E}$-index to analyze the Brazilian passenger air transportation network.
This mode of transport is fundamental for Brazil, since from the 1950s the Brazilian government has discouraged rail passenger transport, making the country highly dependent on road transport~\cite{Railway}. This lack of rail passenger transport linking the different regions and the high cost of maintaining the road network, which leaves many highways in poor condition, in the world's fifth largest country, has created a heavy reliance on air transport for internal mobility in Brazil. Therefore, it is necessary to evaluate the efficiency of the air transport service for the mobility of people over time.

In the first decade of the twentieth century, the Brazilian air market underwent changes in its structure, caused by regulatory policies and other factors. Low-cost companies began operating in the country and large traditional companies closed their activities, increasing the concentration of the air market~\cite{pereira2016metodo}.

Between 2010 and 2012, eight airlines carried more than 99\% of the number of passengers in the country. In 2013, after mergers, that number dropped to six, and since 2014, only four combined companies have already exceeded 99\% of passengers carried~\cite {BrasilAnac}. 
Currently, one of these companies is experiencing financial difficulties and have stopped  its operations, leading to a dispute for their routes among the other companies~\cite{Avianca1}. This indicates an even greater concentration of the Brazilian domestic air market in the short term, which may lead to a reduction in the supply of transportation for the population.

Each airline operates with its own strategic business model. Some have a regional profile, others focus on the demand of large cities, or work with many routes to medium-sized city destinations, among other possibilities. According to data released by the Agência Nacional de Aviação Civil  (ANAC), the number of domestic flight passengers has grown from around 140 billion in 2010 to over 187 billion in 2018~\cite{BrasilAnac}. This means that an increasing number of people have fewer and fewer options for choice of domestic flights in Brazil. The objective of this work is to present the $\mathcal{E}$-index and apply it to investigate the efficiency of the coverage of the Brazilian territory by the air passenger service.

Several papers were published about the Brazilian air transport network by means of statistical mechanics tools and different objectives. L.E.C. Rocha, investigates the structure and evolution of the Brazilian airports network (BAN), between 1995 and 2006, and concluded that it has undergone a strong reconnection in accordance with the changes in demand~\cite {da2009structural}. In addition, the author noted that the size of the network was decreasing (fewer routes), while the number of passengers increased, and this indicated a higher level of occupation of aircraft or larger aircrafts~\cite{da2009structural}. A.V.M. Oliveira \textit{et al.}~\cite{oliveira2016network} analyze the concentration of the Brazilian air transport market before and after the period known as the 2006-2007 aerial "big blackout". This period was marked by many flight delays and cancelations, culminating in two major air crashes in the country. These facts gave rise to the crisis of flight controllers in Brazil, who complained for lack of public investments in the sector. The authors conclude that the Brazilian air transport market was already moving toward concentration even before the aforementioned crisis, but this process accelerated especially at the end of that decade. 

In another sense, S.Wandelt, X. Sun and J. Zhang analyze and compare the evolution of air networks (including the Brazilian one) during the period of 2002 and 2013~\cite {wandelt2019evolution}. They found a decrease in average grade and attributed it to efforts to consolidate the Brazilian air transport business and the privatization of airports, which began in 2010. They also found that the Brazilian network underwent many changes in the period, probably attributed to attempts of reorganization of the airlines after the crisis of air traffic controllers between 2006 and 2007, with repercussion up to 2010.

In the context of the global air transportation network, several works have been carried out in an attempt to analyze the network structure, measuring its complexity. We highlight the work of Sales-Pardo \textit{et al.}~\cite{sales2007extracting}, which analyzes the multilevel modular structure of the global network and T.Verma \textit{et al.}~\cite{verma2014revealing} that verifies the vulnerability of the system through the extraction of network hierarchies. Although both works study network structures in levels, none of them used multiplex approach, which allows us to verify the influence of different airlines in the network and to measure the efficiency of the air service.

We believe that the methodology presented here is useful not only for analyzing air transport networks from another perspective, but also opens the way for quantifying multiplex network efficiency in general, in an unbiased way.

\section*{\centering{Methods}}
\subsection*{Multiplex Efficiency Index ($\mathcal{E}$)}

Next we present the $ \mathcal{E}$-index, which is based on the concept of multiplexed network diversity~\cite{carpi2019assessing}.
Diversity, as a general concept, can be understood as the level of heterogeneity of a system~\cite{weitzman1992diversity,Page:2010:DC:1984802}. In the case of multiplex networks, diversity is related to the variety of connectivity patterns present in the structure. Thus, a greater diversity value implies configurations that exhibit more heterogeneous connectivity patterns within the layers.

Formally, a multiplex network with $ M $ layers, each one having the same set of nodes $ N $, is represented by adjacency matrices  $\mathcal{A}=\{A^{[1]},A^{[2]},\dots,A^{[M]}\}$. From these adjacency matrices it is possible to calculate the distance distribution of nodes (NDD)
~\cite{carpi2019assessing} for each node $i$ in the layer $\overline{p}$ ($\mathcal{N}^{\overline{p}}_i(d)$). It corresponds to the fraction of nodes that are at distance $ d $ (shortest path) of node $ i $ in layer $\overline{p}$; and  the Transition Matrix  ($\mathcal{T}^{\overline{p}}_i(j)$) provides the probability that the node $ j $ in the layer $ \overline{p} $ can be reached, in a single step, by a random walker located in node $i$ in $\overline{p}$. These probability distributions describe the local and global connectivity properties of the $ i $ node in the different layers and can be used to calculate the dissimilarity of the connectivity configurations of the node  $i$ (${\mathcal D_i}$) in layers $\overline{p}$ and $\overline{q}$: ${\mathcal D_i} (\overline{p},\overline{q})=\frac{\sqrt{{\cal J} (\mathcal{N}^{\overline{p}}_i,\mathcal{N}^{\overline{q}}_i)} + \sqrt{{\cal J}  (T^{\overline{p}}_{i},T^{\overline{q}}_{i})}}{2 \sqrt{\log{(2)}}},$ where ${\cal J}$ is the Jensen-Shannon (JS) divergence~\cite{cichocki2010families} that measures the distance between two probability distributions. ${\mathcal D_i}(\overline{p},\overline{q})=0$ if node $i$ has identical connectivity configuration in layers $\overline{p}$ and $\overline{q}$; while ${\mathcal D}_i(\overline{p},\overline{q})=1$, if node $i$ is not connected in one layer, and connected to all nodes in the other.

The average value of ${\mathcal D}_i(\overline{p},\overline{q})$ calculated over all the nodes corresponds to the dissimilarity between the layers $\overline{p}$ and $\overline{q}$, called layer distance (LD):
\begin{equation*}
{\mathcal D}{(\overline{p},\overline{q})}=\langle{\mathcal D}_i(\overline{p},\overline{q})\rangle _i.
\label{LD}
\end{equation*}
If ${\mathcal D}{(\overline{p},\overline{q})}=0$, it implies that the layers $\bar{p}$ and $\bar{q}$ are identical; otherwise, if ${\mathcal D}{(\overline{p},\overline{q})}=1$, one of the layers is fully connected, and the other is totally disconnected.

For completeness, the definition of diversity requires quantifying the distance of a set of elements $S$ for an element $\overline{g}$ that does not belong to $S$, as proposed by Carpi \textit{et al.}~\cite{carpi2019assessing}, should be taken as the minimum distance between the element $S$ for any element in the set $\overline{s_i}$:
\begin{equation}
{\mathcal D}(\overline{g},S)= \min_{\overline{s_i} \in S} {\mathcal D}(\overline{g},\overline{s_i}). 
\label{dist_U}
\end{equation}
The diversity function $U:\tilde{S} \rightarrow \mathbb{R}_{+}$ is defined recursively as 
\begin{equation}
U(S)=max_{\overline{s_i} \in S} \{U(S\setminus{\overline{s_i}}) + \mathcal D(\overline{s_i},S\setminus{\overline{s_i}})\} 
\label{diversity_recurs}
\end{equation}
for all $S \in \tilde{S}$ with $|S| \geq 2$, where $|S|$ represents the cardinality of the set, and $U(S)=0$, for all  $S \in \tilde{S}$  such that $|S|=1$.

The equation~(\ref{diversity_recurs}) shows that the diversity calculation is obtained by summing $M-1$ positive values between $0$ and $1$. 
Thus, the value of $U$ is upper bounded by $M-1$. Thus, it follows that the number of layers considered in the multiplex analysis influences the value of diversity.

The $\mathcal{E}$-index corresponds to a weighted diversity value, avoiding then, the intrinsic bias due to  diversity definition.
$\mathcal{E}$-index is defined as the ratio between the diversity value and the number of layers, as: 
\begin{equation}
\mathcal{E}= \frac{U}{M-1}
\label{diversity_normalized}
\end{equation}
where $0 \leq \mathcal{E} \leq 1$.

Thus, it is possible to compare the $\mathcal{E}$-index between networks without the influence of the number of layers. For networks where there is a large overlap of links, the value of $\mathcal{E}$-index will be close to $0$ (low efficiency). On the other hand, when there is a greater diversification of connections between the layers present in the network, the $\mathcal{E}$-index value will be closer to $1$ (high efficiency). Synthetic examples of the calculation of $\mathcal{E}$-index are presented in Note S1 of the Supplementary Information.

\subsection*{Brazilian air transportation multiplex network (BATMN)}
After the crisis of 2006 and 2007, the Brazilian domestic air network had to reorganize~\cite{wandelt2019evolution,costa2010model}. Since then, a notable increase in domestic demand has been observed in the Brazilian air network (BAN), reaching a maximum during two major events: the 2014 FIFA World Cup and the Rio 2016 Olympic Games~\cite{BrasilAnac}. The Brazilian government invested in the renovation and expansion of airports in the main cities and, as of 2011, started the project of concession of airports for the private sector~\cite{ANACprivatizations}. This scenario led to a major reorganization of BAN, which motivated the present analysis in order to evaluate quantitatively the evolution of the domestic BAN between 2010 and 2018 from the multiplex perspective.

In this way, we constructed the Brazilian air transportation multiplex network (BATMN), through the database available on the ANAC website~\cite {BrasilAnac}. During the study period, the same set of 154 airports was considered each year, representing the network nodes. The links symbolize the existing routes and the layers correspond to the different airlines. As we consider non-directed networks, if a route connecting two airports exists, a link between them is created. According to ANAC, a group of only 8 airlines carried in 2010, more than $99\%$ of passengers, carrying each of them, at least $0.5\%$ of that number. Focusing on the analysis on these companies, we decided to adopt $0.5\%$ of the total number of passengers per year as the lower limit to include an airline as a multiplex network layer.
We constructed monthly networks that correspond to a multiplex structure containing all routes existing for each company in that month, and annual networks that correspond to the multiplex structure considering all the routes traveled for each company throughout the year. In view of these criteria, the diversity measure was computed for an 8-layer network between 2010 and 2012, a 6-layer network in 2013 and a 5-layer network in the following years, until 2018. From this, we calculate the $\mathcal{E}$-index of BATMN. 
 
As Brazil is a member of the International Civil Aviation Organization (ICAO) their airports and airlines are identified by ICAO codes. The Table~\ref{tabelascias}\subref{airlinesICAO} shows the airlines considered in this work and their respective ICAO code, and then the Table~\ref{tabelascias}\subref{airlinesused} shows the companies considered as layers of BATMN, each year.

\begin{table}[ht]
%\caption{Airlines considered in the period}
\begin{subtable}{.48\textwidth}
%\centering
\caption{}
\label{airlinesICAO}
\begin{tabular}{ll}\hline
ICAO Code & Airline Names \\ \hline
AZU & Azul Linhas Aéreas Brasileiras\\
GLO & Gol Linhas Aéreas\\
ONE & Avianca Brasil\\
PTB & Passaredo Transportes Aéreos\\
PTN & Pantanal Linhas Aéreas\\
TAM & TAM Linhas Aéreas\\
TIB & TRIP Linhas Aéreas\\
WEB & Webjet Linhas Aéreas\\ \hline
\end{tabular}
\end{subtable}
\hfill
%\quad
\begin{subtable}{.48\textwidth}
%\centering
\caption{}
\begin{tabular}{ll}\hline
\label{airlinesused}
    Year & Airlines considered \\ \hline
    2010 & AZU, GLO, ONE, PTB, TAM, TIB, PTN, WEB\\
    2011 & AZU, GLO, ONE, PTB, TAM, TIB, PTN, WEB\\
    2012 & AZU, GLO, ONE, PTB, TAM, TIB, PTN, WEB\\
    2013 & AZU, GLO, ONE, PTB, TAM, TIB\\
    2014 & AZU, GLO, ONE, PTB, TAM\\
    2015 & AZU, GLO, ONE, PTB, TAM\\
    2016 & AZU, GLO, ONE, PTB, TAM\\
    2017 & AZU, GLO, ONE, PTB, TAM\\
    2018 & AZU, GLO, ONE, PTB, TAM\\ \hline
\end{tabular}
\end{subtable}
\caption{Airlines considered in this study. The panel~\ref{airlinesICAO} shows the analised airlines and respective ICAO code, and the panel~\ref{airlinesused} displays the airlines of BATMN each year.} 
\label{tabelascias}
\end{table}

\section*{\centering{Results and Discussions}}

\begin{figure}[ht]
  \centering
 \includegraphics[width = 1.0\linewidth]{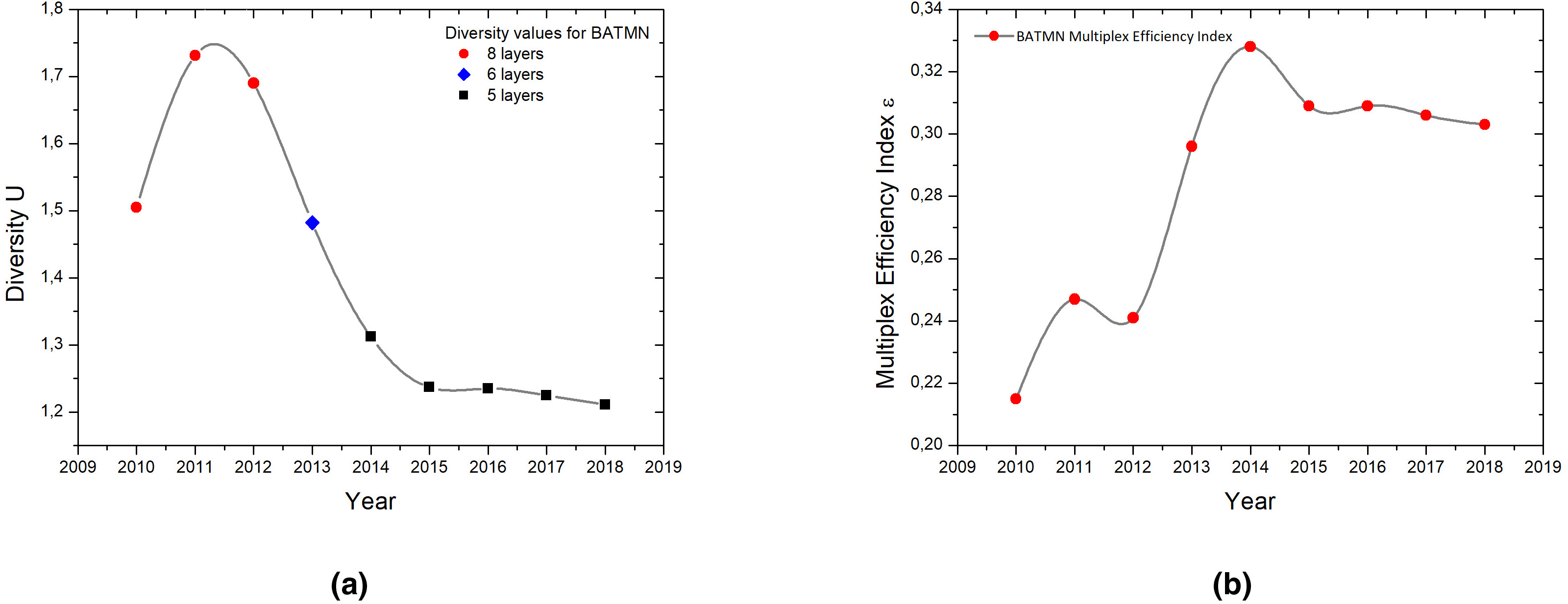}
 \caption{ \label{FigMEI} Analysis of the evolution of BATMN connectivity patterns. \textbf{(a)} Left panel: annual diversity values for BATMN, which are influenced by the number of layers present in the network. \textbf{(b)} Right panel: annual evolution of the $\mathcal{E}$-index for BATMN, which takes into account the number of layers of the network and therefore shows an unbiased result.}
\end{figure}

First, we show the evolution of BATMN's annual diversity values
calculated between 2010 and 2018 - Fig.~\ref{FigMEI}-a. Clearly diversity values strongly correlate with the number of layers in the structure. BATMN is composed of 8 airlines between 2010 and 2012 (red discs), 6 airlines in 2013 (blue disc) and 5 airlines from 2014 to 2018 (black discs). 

On the other hand, $ \mathcal{E}$-index displays a completely different scenario when the number of layers is taken into account by equation~(\ref{diversity_normalized}). Thus, an unbiased analysis of the efficiency of BATMN can be performed, as shown in Fig.~\ref{FigMEI}-b. We can see a local peak of $\mathcal{E}$-index in 2011, followed by a decrease in 2012 and a progressive increase, peaking in 2014. Then, we see a slight drop in $\mathcal{E}$-index in 2015 to a high value when compared to measure before 2014. Since then, $\mathcal{E}$-index has remained stationary around that value.

The $\mathcal{E}$-index is independent of the number of layers, so just reducing layers in BATMN does not lead to an increase in the $\mathcal{E}$-index values. To illustrate this, we show an example in Supplementary Information (Note S1), in which, by reducing a network layer, the value of the $\mathcal{E}$-index decreases.

In analyzing the BAN panorama, it is possible to describe some factors that help explain the observed changes in the $\mathcal{E}$-index values for BATMN. Basically, the merger between companies over the first four years analyzed explains the fluctuations on the efficiency of the network until 2014. Three major merger events were held during this period. In 2010, TAM acquired PTN~\cite{TamPTN} which was a regional airline operating between Congonhas Airport (located in São Paulo) and medium-sized cities.
As of 2011, TAM began to deactivate the PTN routes, until 2013, when the company PTN ceased to operate, with all its routes being incorporated by TAM through ANAC's authorization~\cite{DiarioOf1,TAMdesativatePTN}.
In 2011, GOL acquired $100\%$ of the shares of WEB~\cite{GOLWEB} and, in 2012, consolidated this merger~\cite{GOLWEBCADE}, definitively closing the airline WEB. Finally, AZU merged with TIB in 2012~\cite{TIBAZU} 
and started to control $15\%$ of the domestic market. 
According to the rules established by ANAC, the pre-existing share flights (code-share) between TIB and TAM was gradually shut down by 2014, encouraging competition between AZU and other airlines~\cite{TIBAZUCADE}.

\begin{figure}[ht]
  \centering
 \includegraphics[width = 0.65\linewidth]{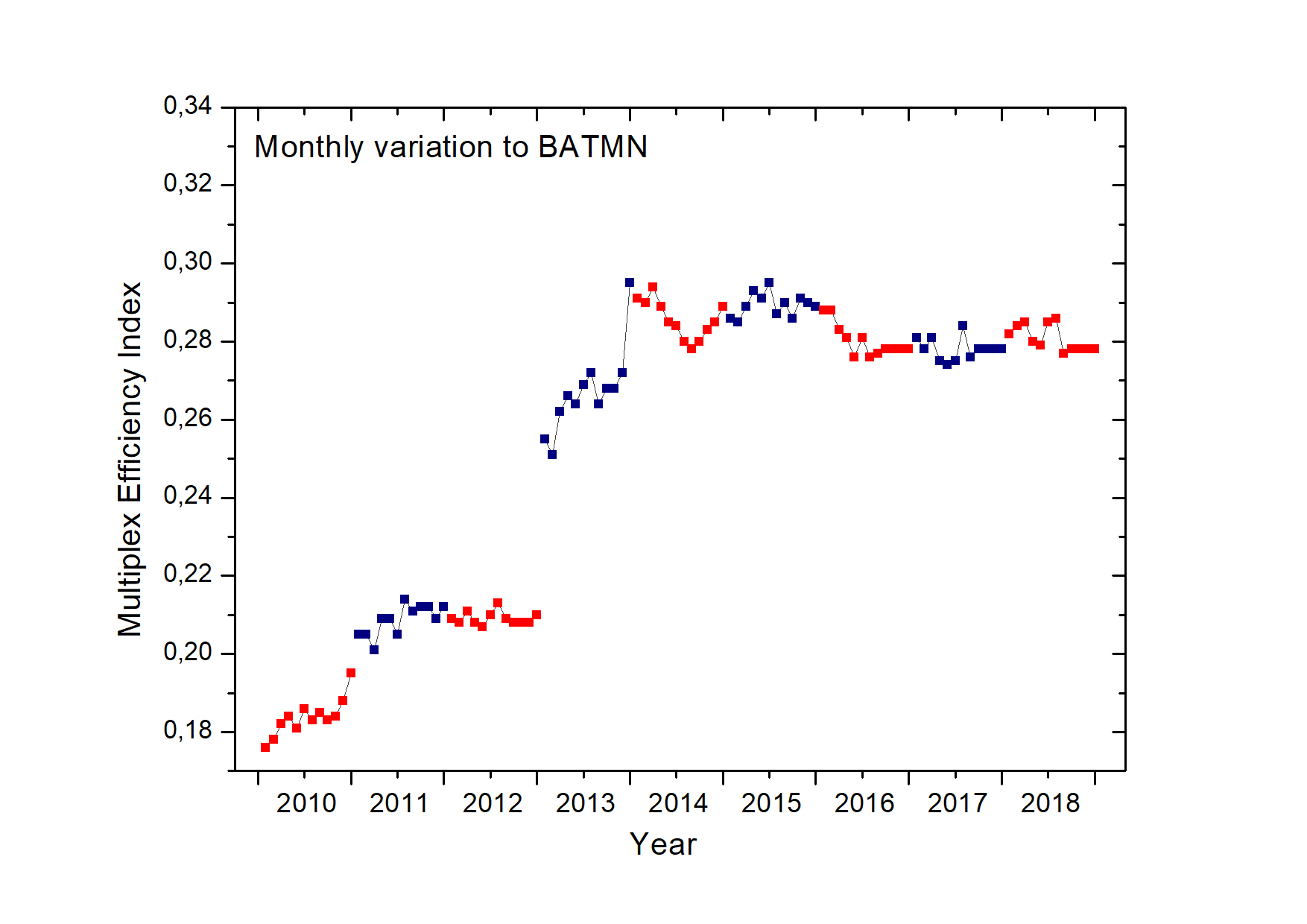}
  \caption{\label{Fig_Montly} Variation of the $\mathcal{E}$-index for BATMN in a monthly network panorama. The dots represent the value of the $\mathcal{E}$-index in consecutive months and alternating colors delimit different years.}

%\dotfill
\end{figure}

To better analyze the effects of these and other events on the network, we perform the monthly calculation of $\mathcal{E}$-index for BATMN, building the network from the routes present in each month, from 2010 to 2018 (see Fig.~\ref{Fig_Montly}). We found a strong jump in the values of $\mathcal{E}$-index from December 2012 to January 2013 and another smaller jump from November to December 2013, which coincides, in the first case, with the exclusion of PTN and WEB from the network and, in the second, excluding TIB. It is also possible to observe in Fig.~\ref{Fig_Montly} that the highest monthly value of $\mathcal{E}$-index corresponds to December 2013 and then there is a stabilization of the $\mathcal{E}$-index, with considerably smaller variations. We realized then that the reorganization of the routes caused by the mergers of the airlines led to an increase in the values of $\mathcal{E}$-index, that is, an increase in the diversification of routes between the participating airlines and, subsequently, the consolidation of this group of companies.

We note that the year 2014 shows an interesting behavior when considering the variations of $\mathcal{E}$-index.
Monthly values show a pronounced minimum, as they decrease from March to August and increase in the following months (Fig.~\ref{Fig_Montly}). This suggests a correlation with the concentration of air routes to meet the growing demand for air transportation to the host cities of the FIFA World Cup games~\cite{AdaptacoesCopa}, held in July. On the other hand, in the annual context, shown in Fig.~\ref{FigMEI}-b, we observe that in 2014 the highest $\mathcal{E}$-index value of the entire period occurs, corresponding to the conclusion of airline mergers and the beginning of a period of stability in the diversification of routes.
%We note that the year 2014 shows an interesting behavior when considering the montly variations of $\mathcal{E}$-index (Fig.~\ref{Fig_Montly}). The minimum pronounced this year coincides with the Fifa World Cup in July, suggesting a correlation with the concentration of routes to meet the growing demand in the host cities~\cite{AdaptacoesCopa}.
%On the other hand, in the annual context, shown in Fig.~\ref{FigMEI}-b, we observe that in 2014 the highest $\mathcal{E}$-index value of the entire period occurs, corresponding to the conclusion of airline mergers and the beginning of a period of stability in the diversification of routes.

\begin{figure}[ht]
  \centering
 \includegraphics[width = 0.6\linewidth]{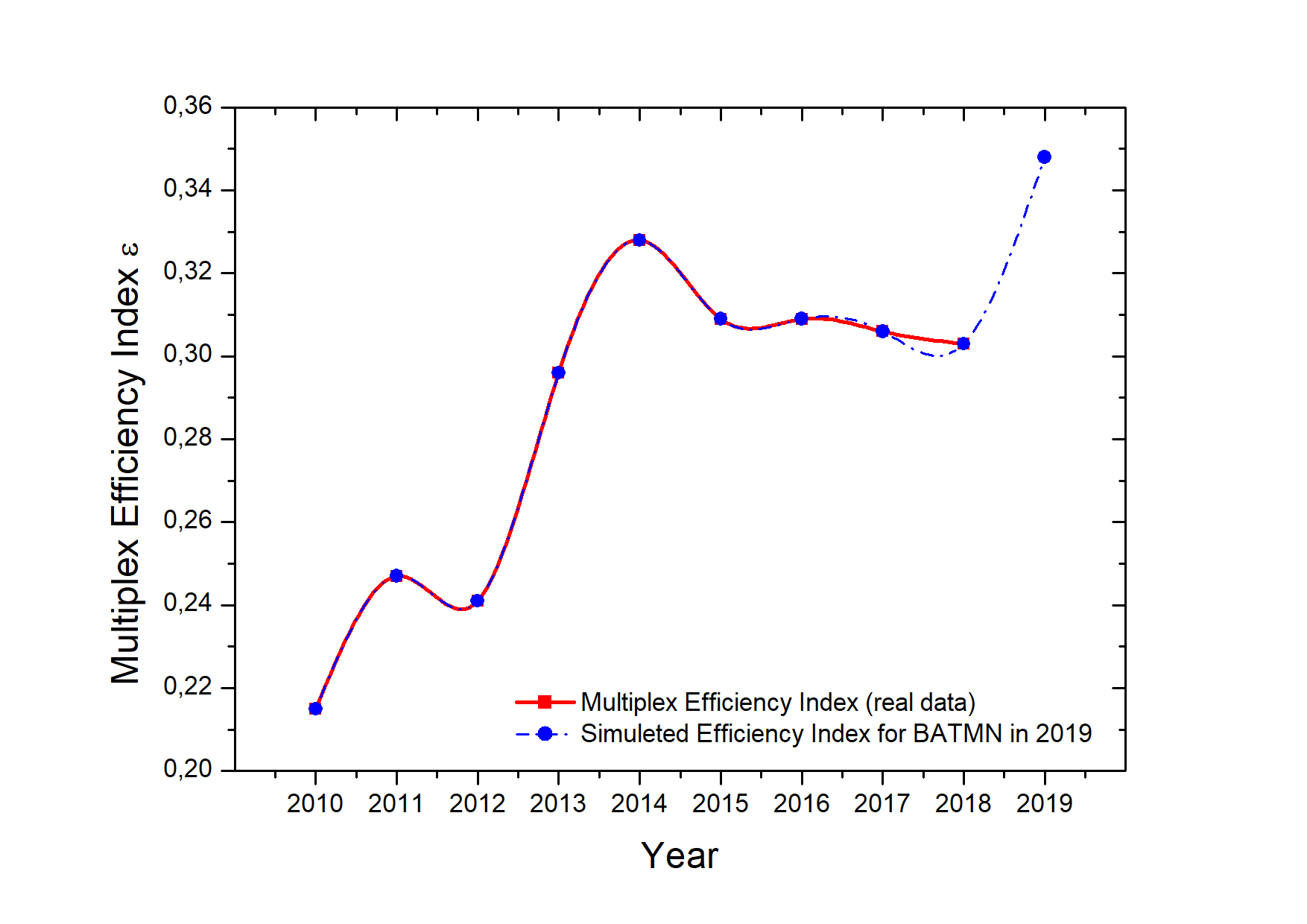}
  \caption{\label{FigMEIlayers} Simulation of the annual $\mathcal{E}$-index values for BATMN in 2019, maintaining the layer configurations of 2018 and excluding the airline ONE. }

%\dotfill
\end{figure}

In yet another reduction in the number of airlines operating in the domestic passenger market in Brazil, airline ONE ended its operations in 2019, following a judicial liquidation process initiated in late 2018~\cite{Aviancarecjud}. As an exercise to estimate the influence of the removal of this layer from BATMN, we simulated the annual network efficiency in 2019 excluding ONE and maintaining the configurations of remaining layers of 2018 - Fig.~\ref{FigMEIlayers}. We can see that the value of $ \mathcal{E}$-index increases after the output of ONE, which means that network coverage is similar, however, with fewer layers. Looking at layer ONE links in 2018, it has 91 routes (BATMN links), of which only 4 are unique to its layer, and therefore $95.6\%$ of routes overlap with the other layers. Thus, the increase of $ \mathcal{E}$-index reflects the exclusion of one layer with high link overlap in relation to the others. This analysis indicates that increased efficiency does not always translate into benefits for the air transport user, as in this case there will be less availability of mobility options.

In order to better understand the influence of each airline on the efficiency of the network, we observe the diversity ordering for each year of the period considered. The diversity ordering indicates the elements in order of their contribution to the diversity of the set, and is obtained by calculating a method based on lexicographic distance~\cite{carpi2019assessing}. The following is the ascending sequence of the BATMN layers according to their contribution to the value of diversity each year.

%\begin{itemize}
\noindent - 2010: ONE, WEB, PTN, TAM, AZU, PTB, GLO, TIB;

\noindent - 2011: WEB, PTN, TAM, ONE, AZU, PTB, GLO, TIB;

\noindent - 2012: PTN, WEB, TAM, ONE, GLO, AZU, PTB, TIB;

\noindent - 2013: TAM, ONE, AZU, GLO, PTB, TIB;

\noindent - 2014: ONE, TAM, GLO, PTB, AZU;

\noindent - 2015: ONE, TAM, GLO, PTB, AZU;

\noindent - 2016: ONE, TAM, GLO, PTB, AZU;

\noindent - 2017: ONE, TAM, GLO, PTB, AZU;

\noindent - 2018: ONE, TAM, GLO, PTB, AZU.

%\end{itemize}

\begin{figure}[htpb]
  \centering
 \includegraphics[width = 1.0\linewidth]{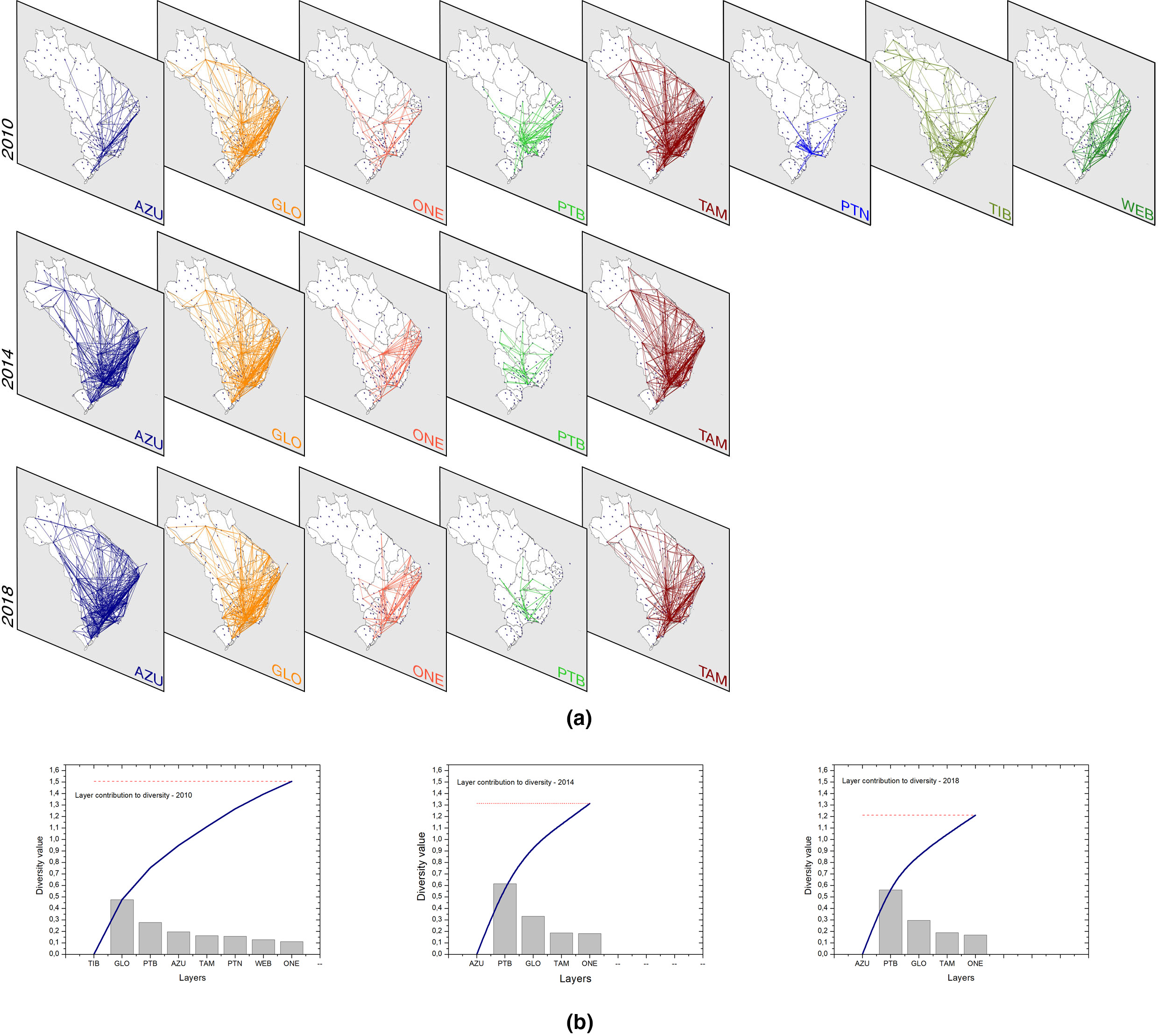}
  \caption{\label{FigOrdering}  BATMN layer evolution for each airline in 2010, 2014 and 2018.
\textbf{(a)} Illustration of BATMN links in each of the three years. \textbf{(b)} Diversity ordering for BATMN in a quantitative way. The columns represent the contribution values of each layer when calculating the network diversity, according to the equation~(\ref{dist_U}). The blue curve shows the cumulative values of these distances at each step of the calculation, as given by equation~(\ref{diversity_recurs}). In the red line we have the final value of the global diversity for BATMN.}
%\dotfill
\end{figure}

Besides that, in Fig.~\ref{FigOrdering}, we show the quantitative participation of each layer in three moments - 2010, 2014 and 2018, respectively, so that Fig.~\ref{FigOrdering}-a shows the BATMN  airline routes (links) for each layer each year. We noticed that PTN and WEB were among the lowest value airlines in calculating diversity, while TIB led this ranking while they were present in the network. These observations are in line with previous discussions made from the analysis of the $\mathcal{E}$-index.

Following the merger between AZU and TIB, AZU became the leader in diversity in 2014 (see Fig.~\ref{FigOrdering}-b). It is worth noting that during the time that the PTN was in the network, there was a progressive decrease of its contribution to the diversity of the multiplex network. Between 2010 and 2012, PTN fell from 6th to 8th position in a group of 8 airlines. This observation coincides with the fact that TAM has reallocated PTN routes to meet its business needs~\cite{TAMdesativatePTN}, causing a decrease in the diversity of the companies. It should be noted that in 2013 the PTN brand still operated, but since it no longer meets the criterion of having more than $ 0.5 \% $ of the total number of passengers, it was not considered in the calculation of diversity.

We see that from 2014 ONE is the layer that contributes least to the diversity of BATMN (see Fig.~\ref{FigOrdering}). Carpi \textit{et al.}~\cite{carpi2019assessing} mention that when a layer is removed, there is a change in diversity that is at least equal to the value of the distance between the corresponding layer and the remaining set, as can be deduced from the equation~(\ref{dist_U}). We then conclude that there is a great overlap of routes between ONE and the other airlines, in agreement with the result already discussed and shown in Fig.~\ref{FigMEIlayers}.

\begin{figure}[ht]
  \centering
 \includegraphics[width = 1.0\linewidth]{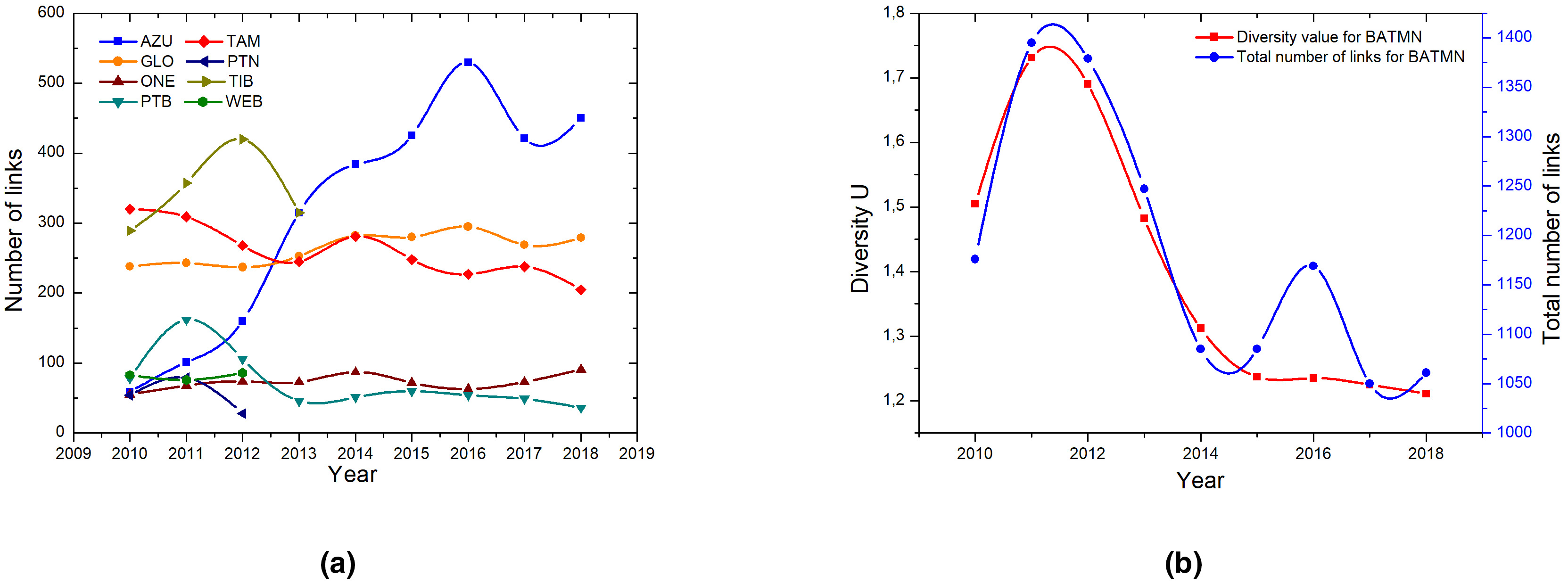}
  \caption{\label{FigRoutes} Analysis of the number of routes (links) of BATMN. \textbf{(a)} Number of links of the airlines studied from 2010 to 2018. The color lines show the number of routes of each airline, specified in the legend. \textbf{(b)} Comparison between the total number of links (blue line) of considered airlines and diversity value for BATMN (red line), for each year from 2010 to 2018.}
%\dotfill
\end{figure}

We investigated the influence of the number of network links on the values of the $\mathcal{E}$-index, analyzing the number of routes (BATMN links), as shown in Fig.~\ref{FigRoutes}-a.
%\sout{The analysis of the number of routes (BATMN links) for the considered airlines is shown in Figure~\ref{FigRoutes}-a.} 
In 2011 and 2012, TIB was the airline with the largest number of routes. In the following year, TIB and AZU led the number of routes. Starting in 2014, after merger between these two airlines and TIB is no longer part of the network, AZU takes the lead with a very significant relative growth. Not surprisingly, the TIB was the leader in ordering diversity by 2013 and, from 2014, AZU is leading the network diversity as shown in Fig.~\ref{FigOrdering}-b.
 
After the merger between TAM and PTN, consolidated in 2013, the growth in the number of TAM routes was not as significant. There was a slight increase in 2014, returning to the same level in 2015, however, this increase may have been influenced by the FIFA World Cup in 2014. Except for 2014, there is a downward trend in the number of TAM routes.

In the third merger in the system, between WEB and GLO, completed in 2012, the number of GLO routes increases from 2013, outperforming its closest competitor, TAM. Both had almost the same number of routes in 2014. Thereafter, the number of GLO routes exceeded that of TAM, increasing further in 2016, when the Rio Olympic Games occurred, returning to the level of 2015 in the following years.

The airline ONE did not show significant changes in the number of routes. It increased in 2014 (FIFA World Cup) and declined in 2016, the year of the Rio Olympic Games. Curiously, its number of routes increased in 2018, the year in which the airline opened the process of judicial administration.

To conclude this analysis, the PTB is the airline that has the lowest number of routes, with a peak in 2011, coinciding with the local peak of the $\mathcal{E}$-index value for BATMN, and falling next. Starting in 2013, its number of routes stabilizes.

We can conclude that, in the case of BATMN, as of 2011, the airline that have a greater number of routes is the one that contribute most to the $ \mathcal{E}$-index of the network, since the surplus of routes cause network diversification.

A remarkable result was obtained by adding the number of routes of the airlines and comparing with the calculation of the diversity. We have verified that there is a high correlation between this number and the value of the network diversity for the BATMN, as shown in Fig.~\ref{FigRoutes}-b. We can observe that a larger number of routes occur when there are lower $\mathcal{E}$-index values. Hence we realize that higher link density on the network leads to a higher probability of link overlap. In 2016, the number of routes increases significantly, probably driven by the Olympic Games in Rio, while the value of diversity and the $\mathcal{E}$-index value increase slightly, which indicates that most new routes correspond to previously existing routes on the network.

All these analyzes demonstrate the utility and importance of the $\mathcal{E}$-index to plan and analyze air mobility in BATMN, as well as in other networks of the same kind.  

In a final survey, we look at the evolution of the local $ \mathcal{E}$-index for the six airports that were privatized before 2017~\cite{privatizations}. The corresponding figures are shown in the Supplementary Information (Note S2). An increase in $\mathcal{E}$-index value for a (local) node means that there is an increase in the connectivity difference between the layers for that node. That is, different airlines serve different cities linked to the same airport. 

Four airports were in the private sector between 2011 and 2012, serving the cities of Natal, Brasília, Campinas and Guarulhos~\cite{privatizations}. The value of $ \mathcal{E}$-index for Campinas (Viracopos International Airport) initially increased and stabilized after 2015. The other airports showed a decrease in the $\mathcal{E}$-index value in the first year after privatization, returning to the previous level and stabilizing in then.

The airports serving the cities of Belo Horizonte (Confins International Airport) and Rio de Janeiro (Galeão International Airport) were privatized in 2013~\cite {privatizations}. Confins airport saw a significant increase in the $ \mathcal{E}$-index in the first year ($ 22 \%$), remaining stable thereafter. Galeão airport increased the value of $ \mathcal{E}$-index in 2014, with inexpressive oscillations, returning to the previous level.

Among the six privatized airports analyzed, four of them tended to stabilize the $\mathcal{E}$-index at a level greater than or equal to the value observed prior to privatization. However, comparing the evolution of $ \mathcal{E}$-index to a major hub in the country (Congonhas Airport~\cite{costa2010model}), which has not yet been privatized, we have seen the same stabilization effect since 2014. Finally, the privatization of airports does not necessarily imply a change in the $\mathcal{E}$-index, but investments in expansion may contribute to this increase, since large investments occurred in the airports of the country to receive the Fifa World Cup of 2014~\cite{investments}.

In a universe of 154 airports considered in this work, the privatization of 6 of them did not imply a significant impact on the BATMN $\mathcal{E}$-index at the global level.

\section*{\centering{Conclusions}}
Combining the $\mathcal{E}$-index and diversity ordering, we have found a way of analyzing the  efficiency of multiplex network, regardless of the number of layers. The $\mathcal{E}$-index allows quantifying the network connectivity heterogeneity  and coverage while the diversity ordering indicates the level of contribution  of each layer, to the diversity of the system.
The greater the difference in connectivity patterns between the layers, the greater the value of the $\mathcal{E}$-index, ranging from 0 to 1.

Applying the $\mathcal{E}$-index to the Brazilian domestic passenger air transport network, we observed that, during the period considered, BATMN exhibited a reorganization of the network in the first four years due to airline mergers. In 2011, regional airlines expanded the number of routes and this fact, combined with the results presented in diversity ordering, explain the local peak observed in annual $\mathcal{E}$-index profile. After all the mergers concluded, the annual value of $\mathcal{E}$-index reaches its maximum in 2014, decreasing slowly to a plateau from 2015. When considering BATMN with monthly temporal precision during the same period, we notice jumps in the values of $\mathcal{E}$-index, which compared to the annual scenario, reflect the fact that the network was reorganizing itself as the new airlines consolidated. The influence of the FIFA World Cup on BATMN is clear when looking at the monthly variation of $\mathcal{E}$-index during 2014, when airlines redirected their routes to meet the demand added by the games.
By simulating the shutdown of one airline in 2019 while maintaining the same outlook as 2018 for other airlines, we see an increase in annual  $\mathcal{E}$-index value. This means that the airline excluded from the network operates a large number of routes overlapping with other airlines. In this context of market concentration, there is no significant reduction of routes, but there is a loss in the offer of airlines to the population.

Finally, we conclude that the privatization of airports does not imply a change in the $\mathcal{E}$-index for the respective airports, but expanding investments to increased efficiency due to increased route offerings. In addition, fluctuations in the $\mathcal{E}$-index of a small fraction of network nodes that do not have a significant impact on the global $\mathcal{E}$-index.

\bibliography{bibliography}

\section*{Acknowledgements}
Research partially supported by Brazilian agencies FAPEMIG, CAPES, and CNPq. LC and IMO thanks for CEFET-MG for financial support. LC thanks CAPES by financial support. APFA thanks to FAPEMIG and CNPq agencies by financial support. APFA thanks CNPq research grant 308792/2018-1.

\section*{Author contributions statement}

I.M.O. planned the experiments, collected data, manipulated data, performed experiments, researched references, interpreted the data and results, constructed figures and tables, and wrote the text. L.C. planned experiments, interpreted data and wrote the text. A.P.F.A. planned experiments, interpreted the results and wrote the text. All authors reviewed the manuscript. 

\section*{Additional information}

\textbf{Supplementary information} accompanies this paper.

\noindent \textbf{Data Availability:} The datasets analysed during the current study are available on the ANAC~\cite {ANACdata} website. 

\noindent \textbf{Competing interests:} The authors declare that there are no competing interests.

\end{document}